\documentclass[twoside]{article}
\usepackage{fleqn,espcrc2}
\usepackage[dvips]{graphicx}

\newcommand{\AmS}{{\protect\the\textfont2
  A\kern-.1667em\lower.5ex\hbox{M}\kern-.125emS}}
\title{DIFFRACTION 2000: New Scaling Laws in Shadow Dynamics}

\author{A.A. Arkhipov\address{Theoretical Physics Division, 
        Institute for High Energy Physics, \\ 
        142284 Protvino, Moscow Region, Russia}}
		
\begin{document}
\def\be{\begin{equation}}					 
\def\ee{\end{equation}}
\def\ber{\begin{eqnarray}}
\def\eer{\end{eqnarray}}	

\begin{abstract}
New scaling structure for the shadow corrections in elastic
scattering from deuteron at high energies has been presented and
discussed. It is shown that this structure corresponds to the
experimental data on proton(antiproton)-deuteron total cross
sections. The effect of weakening for the inelastic screening at
super-high energies has been theoretically predicted.

\rightline{\it\large Perceive the Triangle to see the Truth}
\rightline{Pythagorean School}
\vspace{0.3pc}
\end{abstract}

\maketitle


 
\section{INTRODUCTION}

Experimental and theoretical studies of high-energy particle
interaction with deuterons have shown that total cross section in
scattering from deuteron cannot be treated as equal to the sum of
total cross sections in scattering from free proton and neutron even
in the range of asymptotically high energies. Glauber  was the first
who proposed an explanation of this effect. Using the methods of
diffraction theory, quasiclassical picture for scattering from
composite systems and eikonal approximation for high-energy
scattering amplitudes he had found long ago  \cite{1} that total
cross section in scattering  from deuteron can  be expressed with the
formula
\be
\sigma_d = \sigma_p +\sigma_n - \delta\sigma, \label{1}
\ee
where
\be
\delta\sigma = \delta\sigma_G =
\frac{\sigma_p\cdot \sigma_n}{4\pi}<\frac{1}{r^2}>_d.\label{2}
\ee
Here $\sigma_d, \sigma_p, \sigma_n$ are the total cross sections
in scattering from deuteron, proton and neutron, $<r^{-2}>_d$ is the
average value for the inverse square of the distance between the
nucleons (separation of the nucleons) inside of a deuteron, $\delta
\sigma_G$ is the Glauber shadow  correction describing the effect of
eclipsing or the screening effect in the recent terminology.
The Glauber shadow correction  has a quite clear physical
interpretation. This correction originates from elastic rescattering
of an incident particle on the nucleons in a deuteron and corresponds
to the configuration when the relative position of the nucleons in a
deuteron is such that one casts its ``shadow" on the other \cite{1}.

It was soon understood that in the range of high energies the shadow
effects may arise due to inelastic interactions of incident particle
with the nucleons of deuteron \cite{2,3,4,5,6}. Therefore inelastic
shadow correction had to be added to the Glauber one.

A simple formula for the total (elastic plus inelastic) shadow
correction had been derived by Gribov \cite{4} in the assumption of
Pomeron dominance in the dynamics of elastic and inelastic
interactions. However it was observed that the calculations performed
by Gribov formula did not met the experimental data: the calculated
values of the inelastic shadow correction over-estimated the
experimental values. 

The idea that the Pomeron dominance is not justified at most at the
recently available energies has been explored in the papers \cite{5}.
The authors of Refs \cite{5} argued that account of the
triple-reggeon diagrams for six-point amplitude in addition to the
triple-pomeron ones allowed them to obtain a good agreement with the
experiment. Alberi and Baldracchini replied \cite{6} and pointed out
that discrepancy between theory and experiment cannot be filled by
taking into account the triple-reggeon diagrams: in fact, it needed
to modify  the dynamics of the six-point amplitude with more
complicated diagrams than the triple Regge ones. This means that up
to now we have not in the framework of Regge phenomenology a clear
understanding for the shadow corrections in elastic scattering from
deuteron.

The theoretical understanding of the screening effects in scattering
from any composite system has the fundamental importance, because the
structure of shadow corrections is deeply related to the structure of
composite system as itself. At the same time the shadow corrections
structure displays the new sites of the fundamental dynamics.

Here we are concerned with the study of sha\-dow dynamics in
scattering
from deuteron in some details. The new scaling characteristics of
scattering from the constituents in the composite systems, having a
clear physical interpretation, are established.

\section{SCATTERING FROM DEUTERON, THREE-BODY FORCES AND SING\-LE
DIFFRACTION DISSOCIATION}

In our papers \cite{7,8,9} the problem of scattering from two-body
bound states was treated with the help of dynamic equations obtained
on the basis of single-time formalism in QFT \cite{7}. As has been
shown in \cite{8,9}, the total cross-section in the scattering from
deuteron can  be expressed by the formula
\be
\sigma_{hd}^{tot}(s) = \sigma_{hp}^{tot}(\hat s)
+\sigma_{hn}^{tot}(\hat s) -
\delta\sigma(s), \label{3}
\ee
where $\sigma_{hd}, \sigma_{hp}, \sigma_{hn}$  are the total
cross-sections in scattering from deuteron, proton and neutron, 
\begin{equation}
\delta\sigma(s) = \delta\sigma^{el}(s)
+\delta\sigma^{inel}(s),\label{4}
\end{equation}
\begin{equation}
\delta\sigma^{el}(s) =
\frac{\sigma_{hp}^{tot}(\hat s) \sigma_{hn}^{tot}(\hat s)}{4\pi(
R^2_d+B_{hp}(\hat s)+B_{hn}(\hat s))},\, \hat s = \frac{s}{2},
\label{5}
\end{equation}
$B_{hN}(s)$ is the slope of the forward diffraction peak in the
elastic scattering from nucleon, $1/R_d^2$ is defined by the deuteron
relativistic formfactor \cite{8}, $\delta\sigma^{el}$ is the shadow
correction describing the effect of eclipsing or screening effect
during elastic rescatterings of an incident hadron on the nucleons in
a deuteron.

The quantity $\delta\sigma^{inel}$ in our approach represents the
contribution of the three-body forces to the total cross-section in
the scattering from deuteron. The definition of three-body forces in
relativistic quantum theory see the recent paper \cite{10} and
references therein. 

For simplicity, we consider the model proposed in \cite{9} where the
imaginary part of the three-body forces scattering amplitude has the
form 
\ber
\lefteqn{Im\,{\cal F}_0(s; \vec p_1, \vec p_2, \vec p_3; \vec q_1,
\vec q_2,
\vec q_3) =}  \nonumber\\
& & f_0(s) \exp \Biggl\{-\frac{R^2_0(s)}{4} \sum^{3}_{i=1} (\vec
p_i-\vec
q_i)^2\Biggr\},\label{9}
\eer
This model assumption is not so significant for our main conclusions
but allows one to make some calculations in a closed form

From the analysis of the problem of high-energy particle
scattering from deuteron we have derived the formula relating
one-particle inclusive cross-section with the imaginary part of the
three-body forces scattering amplitude (see details in \cite{10} and
references therein). In this way we can establish
a deep connection of inelastic shadow correction with one-particle
inclusive cross-section which allows one to express the
inelastic shadow correction via total single diffractive dissociation
cross-section. In fact, let us define the total single diffractive
dissociation cross-section by the formula
\be
\sigma_{sd}^{\varepsilon}(s) =
\pi\int_{M_{min}^2}^{\varepsilon s}\frac{dM_X^2}{s}
\int_{t_{-}(M_X^2)}^{t_{+}(M_X^2)} dt \frac{d\sigma}{dtdM_X^2}.
\label{22}
\ee
Here we have specially labeled the total single diffractive
dissociation cross-section by the index $\varepsilon$. It's clear the
parameter $\varepsilon$ defines the range of integration in the
variable $M_X^2$. Unfortunately up today there is no common consent
in the choice of this parameter. However we would like to point out
the exceptional value for the parameter $\varepsilon$ which naturally
arises from our approach. Namely, let us put $\varepsilon^{ex} =
\sqrt{2\pi}/2M_N R_d$, then we  define the exceptional total single
diffractive dissociation cross-section $\sigma_{sd}^{ex}(s) =
\sigma_{sd}^{\varepsilon}(s)|_{\varepsilon=\varepsilon^{ex}}$.
As a result we obtain 
\be
\delta\sigma^{inel}(s) = 2
\sigma_{sd}^{ex}(s)a^{inel}(x_{inel}),\label{23}
\ee
where
\be
a^{inel}(x_{inel}) = \frac{x_{inel}}{(1+x_{inel})^{3/2}},\label{24}
\ee
\be
x_{inel} \equiv \frac{R_0^2(s)}{R_d^2} =
\frac{2B_{sd}(s)}{R_d^2}.
\ee
See the definition of $B_{sd}$ in \cite{11}.

Here is just the place to rewrite the elastic shadow correction
(\ref{5}) in a similar form
\be
\delta\sigma^{el}(s) = 2 \sigma^{el}(s)a^{el}(x_{el}), \label{25}
\ee
where
\be
a^{el}(x_{el}) = \frac{x_{el}}{1+x_{el}},\label{26}
\ee
\be
x_{el} \equiv \frac{2B_{hN}^{el}(s)}{R_d^2} =
\frac{R_{hN}^2(s)}{R_d^2}.
\ee

The obtained expressions for the shadow corrections have a quite
transparent physical meaning, both the elastic $a^{el}$ and
inelastic $a^{inel}$ scaling functions have a clear physical
interpretation. The function $a^{el}$  measures out a portion of
elastic rescattering events among of all events during  an
interaction of an incident particle with a deuteron as a whole, and
this function attached to the total probability of elastic
interaction of an incident particle with a separate nucleon in a
deuteron. Correspondingly the function $a^{inel}$ measures out a
portion of inelastic events of inclusive type among of all events
during  an interaction of an incident particle with a deuteron as a
whole, and this function attached to the total probability of
single diffraction dissociation  of an incident particle on a
separate nucleon in a deuteron. The scaling variables $x_{el}$ and
$x_{inel}$ have a quite clear physical meaning too. The dimensionless
quantity  $x_{el}$ characterizes the effective distances measured in
the units of ``fundamental length", which the deuteron size is, in
elastic interactions, but the similar quantity $x_{inel}$
characterizes the effective distances measured in the units of the
same ``fundamental length" during inelastic interactions.

The functions $a^{el}$ and $a^{inel}$ have a quite different
behaviour: $a^{el}$ is a monotonic function while $a^{inel}$ has a
maximum at the point $x^{max}_{inel}=2$ where
$a^{inel}(x^{max}_{inel})=2/3\sqrt{3}$. This fact results an
interesting physical effect of weakening of the inelastic eclipsing
(screening) at superhigh energies. The energy at the maximum of
$a^{inel}$ can easily be calculated from the equation $R_0^2(s)=2
R_d^2$.

Account of the real part for the hadron-nucleon elastic scattering
amplitude  modifies the scaling function $a^{el}$ in the following
way
\be
a^{el}(x_{el}) \longrightarrow
a^{el}(x_{el},\rho_{el})=a^{el}(x_{el})\frac{1-\rho_{el}^2}{1+
\rho_{el}^2}.
\ee
The scaling function $a^{inel}$ is not modifying.

\section{COMPARISON TO THE EXPERIME\-NTAL DATA}

We have tried to make a preliminary comparison of the new structure
for the shadow corrections in elastic scattering from deuteron to the
existing experimental data on proton-deuteron and antiproton-deuteron
total cross sections. 

In the first step we analysed the experimental data on
antiproton--deute\-ron total cross sections. We have used  our
theoretical formula describing the global structure of
antiproton-proton total cross sections \cite{10,12} as $\sigma_{\bar
p p}^{tot}=\sigma_{\bar p n}^{tot}\equiv\sigma_{\bar p N}^{tot}$. The
new fit to the data on the total single diffraction dissociation
cross sections in $\bar p p$ collision with the formula from
\cite{10} has been made as well using a wider set of the data (see
\cite{13,14,15,16,17,18}), $R_d^2$ was considered as a single
free fit parameter. Our fit yielded $R_d^2 = 66.61 \pm 1.16
\,GeV^{-2}$. The fit result is shown in Fig. 1.
\begin{figure}[htb]
\vspace{20pt}
\begin{picture}(150,82)
\put(15,-5){\includegraphics[scale=0.65]{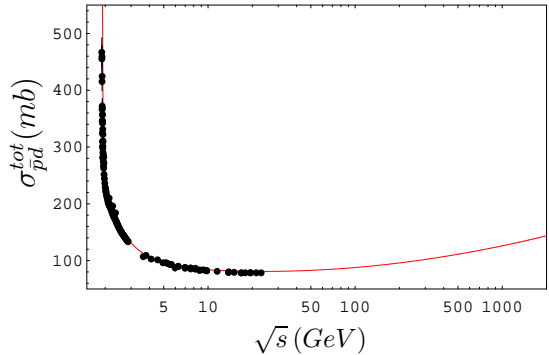}}
\put(92,-15){$\sqrt{s}\, (GeV)$}
\put(0,45){\rotatebox{90}{\large$\sigma^{tot}_{\bar{p}d} (mb)$}}
\end{picture}
\caption{The total antiproton-deuteron cross-section versus
$\sqrt{s}$ compared with the theory. Statistical and systematic
errors added in quadrature.}
\label{fig:1}
\vspace{-15pt}
\end{figure}
The fitted value for the $R_d^2$ satisfies with a good accuracy to
the equality $R_d^2 = 2/3\,r^2_{d,m}\, (r_{d,m}=1.963(4)fm$
\cite{19}).

After that it was very intriguing for us to make a comparison to the
data on proton-deuteron total cross sections where  $R_d^2$ has to be
fixed from the previous fit to the data on antiproton-deuteron total
cross sections. As in the previous fit we supposed
$\sigma_{pp}^{tot}=\sigma_{pn}^{tot}\equiv\sigma_{pN}^{tot}$ and
$\sigma_{pp}^{tot}$ had been taken from our global description of
proton-proton total cross sections \cite{10,12}. We also assumed that
$B_{pN}^{el}=B_{\bar p N}^{el}$. So, in this case we have not any
free parameters. The result of the comparison is shown  in Fig. 2.
\begin{figure}[htb]
\vspace{20pt}
\begin{picture}(144,82)
\put(15,-5){\includegraphics[scale=0.65]{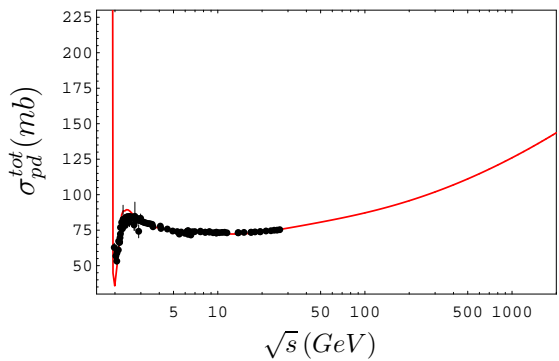}}
\put(92,-15){$\sqrt{s}\, (GeV)$}
\put(-4,44){\rotatebox{90}{\large$\sigma^{tot}_{pd} (mb)$}}
\end{picture}
\caption{The total proton-deuteron cross-section versus
$\sqrt{s}$ compared with the theory without any free parameters.
Statistical and systematic errors added in quadrature.}
\label{fig:2}
\vspace{-15pt}
\end{figure}
As you can see the correspondence of the theory to the experimental
data is quite remarkable apart from the resonance region. The
resonance region requires a more careful consideration than that
performed here.

\section{SUMMARY AND DISCUSSION}

In this report we have been concerned with a study of shadow
corrections to the total cross section in scattering from deuteron.
The dynamic apparatus based on the single-time formalism in QFT has
been used  as a tool and subsequently applied to describe the
properties of high-energy particle interaction in scattering from
two-body composite system. In this way we found the new structures
for the total shadow correction to the total cross section in
scattering from deuteron. 

First of all, it was observed that the total shadow correction
inherits the general structure of total cross section and contains
two
inherent parts as well, elastic part and inelastic one. This
partitioning has been performed explicitly in the framework of our
approach. It turns out that the elastic part can be expressed
through the elastic scaling (structure) function and the fundamental
dynamical quantity, which is the total elastic cross section in
scattering from an isolated constituent (nucleon) in the composite
system (deuteron). At the same time the inelastic part is expressed
through the inelastic scaling (structure) function and the
fundamental
dynamical quantity, which is the total single diffractive
dissociation cross section in scattering from an isolated constituent
in the composite system too. Thus the general formalism in QFT makes
it possible to define properly the dynamics of particle scattering
from composite system and express this dynamics in terms of the
fundamental dynamics of particle scattering from an isolated
constituent in the composite system and the structure of the
composite system as itself.

What was probably the most important, which we have discovered in the
work, elastic and inelastic structure functions have a quite
different behaviour. The inelastic structure function has a maximum
and tends to zero at infinity, while the elastic structure function
is a monotonic function and tends to unity at infinity. This is the
most significant difference between the elastic and inelastic
structure functions and it has far reaching physical consequences.
This difference manifests itself in the effect of weakening of
inelastic eclipsing (screening) at superhigh energies. What does
it mean physically? To understand it let's combine the elastic shadow
correction and the first terms in Eq. (\ref{3}) for the
hadron-deuteron total cross section 
\be
\sigma_{hd}^{tot} = 2 \sigma_{hN}^{inel} + 2 \sigma_{hN}^{el}(1
-
a^{el}) - \delta\sigma^{inel}. \label{34}
\ee
We have in this way that asymptotically
\be
\sigma_{hd}^{tot} = 2 \sigma_{hN}^{inel},\quad s\longrightarrow
\infty.
\ee
Probably the generalization of this result to any many-nucleon
systems (nuclei) looks like
\be
\sigma_{hA}^{tot} = A \sigma_{hN}^{inel},\quad s\longrightarrow
\infty.
\ee
Besides, we would like to emphasize the different range of variation
for the elastic and inelastic structure functions
\be
0 \leq a^{el} \leq 1,\qquad 0 \leq a^{inel} \leq 2/3\sqrt{3}.
\ee

The energy, where the inelastic shadow correction has a maximum, has
to be calculated from the equality $R_0^2(s_m)=2R_d^2$. Taking
$R_d^2=66.61$ from the fit and $R_0^2(s)$ from the paper \cite{12},
we obtain $\sqrt{s_m}=9.01\,10^8\,GeV=901\,PeV$. Of course such
energies are not available at the recently working accelerators.
However, we always have a room for a speculative discussion. For
example, let us consider the proton as a two-body (quark-diquark)
composite system. From the experiment it is known the value for the
charge radius of the proton $r_{p,ch}=0.88\,fm$. If we put
$R_p^2=2/3\,r^2_{p,ch}$, then resolving the equation
$R_0^2(s_p)=R_p^2$ we obtain $\sqrt{s_p}=1681\,GeV$. This is just the
energy of Tevatron. 

Our analysis shows that the magnitude of inelastic shadow correction
is about 10 percent from elastic one at recently available energies.
That is why the precise measurements of hadron-deuteron total
cross-sections at high energies are the most important. 

\section*{ACKNOWLEDGEMENTS}
It is my great pleasure to express thanks to the Organizing 
Committee for the kind invitation to attend the Workshop
``Diffraction 2000". I would like especially to thank Roberto Fiore,
Laszlo Jenkovsky, Alessandro Papa, Franca Morrone and Maxim Kotsky
for warm and kind hospitality throughout the Workshop.

\end{document}